\documentclass[fleqn,twoside]{article}

\usepackage{amsmath}
\usepackage{amsfonts}

\usepackage[headings]{espcrc2}

\readRCS
$Id: espcrc2.tex,v 1.2 2004/02/24 11:22:11 spepping Exp $
\usepackage{graphicx}
\usepackage[figuresright]{rotating}

\newcommand{\Lie}[1]{\mathcal{#1}}

\newcommand{\td}{\mathrm{d}}

\title{A note on backreacting flavors from calibrated geometry}

\author{Johannes Schmude\address{
  Department of Physics\\
  Swansea University, Swansea, SA2 8PP, United Kingdom}%
  \thanks{I would like to thank J\'er\^ome Gaillard anc Carlos
    N\'u\~nez for comments on the manuscript. My work is funded by the
    German National Academic Foundation (Studienstiftung des deutschen
    Volkes) as well as an STFC studentship.}
}
       
\runtitle{A note on backrecting flavors from calibrated geometry}
\runauthor{J. Schmude}

\begin{document}

\begin{abstract}
  One of the main problems in the search for string duals with
  backreacting, smeared flavors is the construction of a suitable
  source density. We review how this issue may be addressed using
  generalized calibrated geometry.
  \vspace{1pc}
\end{abstract}

\maketitle

\section{Introduction}

Gauge/string duality in its original formulation §
\cite{Maldacena:1997re,Witten:1998qj} relates strongly coupled
$d=3+1$, $\mathcal{N}=4$ super Yang-Mills to weakly coupled type IIB
string theory on $AdS_5 \times S^5$ and vice-versa. It
did not take long for the duality to be generalized to gauge theories
in different dimension \cite{Itzhaki:1998dd} or with less
supersymmetry \cite{Klebanov:1998hh,Maldacena:2000yy}. In all these
cases, the dynamics of the gauge theory are captured by a closed string
theory on a suitable ten-dimensional space-time.

If one adds an open string sector by the inclusion of D-branes, one
introduces fields into the gauge theory that transform under the
fundamental representation of the gauge group \cite{Karch:2002sh}. If
these branes extend along a non-compact cycle transverse to the
$\mathbb{R}^{1,d-1}$ associated with the gauge theory, the local
$\Lie{SU}(N_f)$ gauge symmetry living on these branes turns into a
global flavor symmetry of the dual gauge theory -- one has flavored
the theory.

For many purposes, it is sufficient to
work in the limit $N_f \ll N_c$, in which the backreaction of the
branes onto the geometry may be ignored. In the perturbative regime of
the gauge theory this corresponds to the exclusion of flavor loops
from all Feynman diagrams. It may be conceptually straightforward
to go beyond this approximation and find duals for $N_f \sim N_c$
by including the backreaction of the flavor branes onto the
geometry; yet it should be no surprise that the technical challenges
in doing so are often quite formidable. The issue was first addressed
in \cite{Burrington:2004id,Kirsch:2005uy}. One such complication
is the fact that localized branes add delta-function sources to the
equations of motion. The standard method of dealing with these relies
on smearing the flavor branes over their transverse directions
\cite{Casero:2006pt}, turning the delta-functions into smooth source
distributions. As was shown in \cite{Gaillard:2008wt}, the
construction of the source-densities involved can be helped by making
use of generalized calibrated geometry \cite{Gutowski:1999tu}.
In this note we shall review the construction of flavored supergravity
duals and show what can be learned from generalized calibrated
geometry.\footnote{For a more complete list of references to the
  subject see references to \cite{Casero:2006pt} and those in
  \cite{Gaillard:2008wt}.}

\section{Flavored supergravity duals from calibrated geometry}
\label{sec:flavored-supergravity-duals}

Simply said, a string dual in the supergravity limit consists of the
metric ($g_{mn}$) the dilaton ($\Phi$) and a set of RR ($F_{(i)}$) or NS
($H_{(3)}$) gauge fields, whose dynamics are captured by the relevant type
IIA/B action $S_{\text{IIA/B}}$. On an intuitive level, it is clear
that the addition of backreacting sources will deform the geometry of
the background, yet conserve its essential topological
features. Therefore one usually begins the flavoring procedure by
studying deformations of the original supergravity dual. This gives a
suitable ansatz for the geometry of the flavored background. Then, in
order to add backreacting Dp-brane sources to the system, one
considers the combined action
\begin{equation}
  \label{eq:combined_action}
  S = S_{\text{IIA/B}} + S_{\text{branes}}
\end{equation}
where the brane action $S_{\text{branes}}$ is given by the usual DBI
and WZ terms. The presence of $S_{\text{branes}}$ adds sources
to the type IIA/B equations of motion. Our main interest
lies in the resulting violation of the Bianchi identity\footnote{We
  shall work in String frame.} $\td F_{(10-p-2)} = 0$,
\begin{equation}
  \label{eq:modified_bianchi}
  \td F_{(10-p-2)} = 2 \kappa_{10}^2 T_p \Omega
\end{equation}
$\Omega$ plays the role of a source density. For localized branes it
consists of isolated delta functions, which are replaced by a
continuous charge density upon smearing. Its construction is often one
of the major problems encountered when flavoring a supergravity dual,
as it has to encode the brane embeddings as well as their distribution
over cycles transverse to their world volumes. We will see soon how
this issue can be simplified by using calibrated
geometry.\footnote{See however \cite{Bigazzi:2008zt}.} Subsequently,
one looks for solutions of the modified equations of motion using the
deformed ansatz.

As usual, supersymmetry makes things a great deal easier, because
second order equations of motion can be traded for first order BPS
ones. One should note that the latter are also modified by the
presence of the source term. Due to a theorem by Koerber and Tsimpis
\cite{Koerber:2007hd}, solutions of the modified BPS
equations satisfying (\ref{eq:modified_bianchi}) are solutions of the
full modified equations of motion.

If we want to use the BPS equations as an aid towards constructing a
background, the flavor brane embeddings $X(\xi)$ have to preserve some
of the supersymmetries of the background. In the flavoring literature,
the standard tool for discussing supersymmetric brane embeddings is
$\kappa$-symmetry \cite{Bergshoeff:1996tu}. Let $\epsilon$ be a SUSY
spinor of the background. For a brane embedding $X(\xi)$, one
constructs the $\kappa$-symmetry matrix $\Gamma_\kappa[X(\xi)]$, that
acts on the spinors of the background. The embedding is supersymmetric
if
\begin{equation}\label{eq:kappa-symmetry_condition}
  \Gamma_\kappa[X(\xi)] \epsilon = \epsilon  
\end{equation}
This condition can be rephrased using generalized calibrated geometry
\cite{Gutowski:1999tu,Koerber:2005qi,Martucci:2005ht,Koerber:2006hh}. Here one defines a $p+1$-form, the calibration
form, along the lines of
\begin{equation}
  \label{eq:calibration_form}
  \hat{\phi} = \frac{1}{(p+1)!} ( \epsilon^\dagger \Gamma_{a_0 \dots a_p}
  \epsilon ) e^{a^0 \dots a^p}
\end{equation}
Supersymmetry is then satisfied if the volume form induced onto the
cycle defined by $X(\xi)$ equals the pull-back of the calibration:
\begin{equation}
  \label{eq:calibration-condition}
  X^*\hat{\phi} = \sqrt{-g_{\text{ind}}} \td^{p+1}\xi
\end{equation}
It is quite crucial that, for fixed $p$, the calibration form
$\hat{\phi}$ is the same for all Dp-branes, independent of their
embedding, while $\Gamma_\kappa[X(\xi)]$ is not. In other words,
$\hat{\phi}$ knows about all SUSY embeddings of the background.

For backgrounds without fluxes, supersymmetric branes are known to
wrap minimal volume cycles. This fact reappears in calibrated geometry
as the closure of the calibration $\td \hat{\phi} = 0$. Fluxes deform
such embeddings, and so the supersymmetry condition becomes
\begin{equation}
  \td (e^{-\Phi} \hat{\phi}) = F_{(p+2)}
\end{equation}
If one combines this with the modified Bianchi identity
(\ref{eq:modified_bianchi}), one arrives at
\begin{equation}
  \label{eq:smearing_from_calibration}
  \td [ * \td (e^{-\Phi}\hat{\phi})] = 2\kappa_{10}^2 T_p \Omega
\end{equation}
One should note, that this relation gives strong constraints on the
smearing form -- an issue that was first exploited in the context of
flavored duals in \cite{Gaillard:2008wt} to show that the source
distribution form has to respect certain symmetries of the background
-- yet does not fix it. The calibration depends on the vielbein $e^a$,
which again depends on the original deformed ansatz. However, as the
calibration form captures the embeddings of all possible
supersymmetric Dp-branes, the smearing form in its general form of
(\ref{eq:smearing_from_calibration}) knows about all possible ways of
smearing them. One can constrain the original ansatz for the flavored
background from the knowledge of the general structure of these two
differential forms.

\end{document}